\documentclass[]{spie}  %>>> use for US letter paper
\usepackage[]{graphicx}

\title{Simulation of Êa method to directly image  exoplanets around multiple stars systems.Ê} 

\author{Sandrine J. Thomas\supit{a}, Eduardo Bendek\supit{b}, Ruslan Belikov\supit{b},
\skiplinehalf
\supit{a} UARC/NASA Ames Research Center, Moffett Field, CA, USA; \\
\supit{b} NASA Ames Research Center, Moffett Field, CA, USA
}

\authorinfo{Further author information: (Send correspondence to Sandrine Thomas)\\Sandrine Thomas: E-mail: sandrine.j.thomas@nasa.gov, Telephone: 1 650-604-4297}
\begin{document} 
  \maketitle 

%%%%%%%%%%%%%%%%%%%%%%%%%%%%%%%%%%%%%%%%%%%%%%%%%%%%%%%%%%%%% 
\begin{abstract}
Abstract: Direct imaging of extra-solar planets has now become a reality, especially with the deployment and commissioning of the first generation of specialized ground-based instruments such as the GPI, SPHERE, P1640 and SCExAO. These systems will allow detection of planets $10{^7}$ times fainter than their host star.  
Ê
For space-based missions, such as EXCEDE, EXO-C, EXO-S, WFIRST/AFTA,Êdifferent teams have shown in laboratories contrasts reaching $10^{-10}$ within a few diffraction limits from the star using a combination of a coronagraph to suppress light coming from the host star and a wavefront control system. These demonstrations use a deformable mirror (DM) to remove residual starlight (speckles) created by the imperfections of telescope. However, all these current and future systems focus on detecting faint planets around a single host star or unresolved binaries/multiples, while several  targets or planet candidates are located around nearby binary stars such as our neighbor star Alpha Centauri. 
Ê
Until now, it has been thought that removing the light of a companion star is impossible with current technology, excluding binary star systems from target lists of direct imaging missions. Direct imaging around binaries/multiple systems at a level of contrast allowing Earth-like planet detection is challenging because the region of interest, where a dark zone is essential, is contaminated by the light coming from the hostÕs star companion. We propose a method to simultaneously correct aberrations and diffraction of light coming from the target star as well as its companion star in order to reveal planets orbiting the target star. This method works even if the companion star is outside the control region of the DM (beyond its half-Nyquist frequency), by taking advantage of aliasing effects. 
\end{abstract}

%>>>> Include a list of keywords after the abstract 

\section{Introduction}

The field of exoplanets search is rapidly expending with the success of the Kepler mission \cite{Burke14} (and reference therein) and the emergence of direct imaging ground based instruments (GPI\cite{Macintosh14}, SPHERE\cite{Beuzit08}, SCExAO\cite{Guyon10}, P1640\cite{Hinkley08}). The ultimate goal of these missions is to seek other Earths in the Universe and detect life on them. The Kepler space telescope has already revealed that roughly 22\% of stars have planets close to Earth size in regions close to their habitable zones. The next natural step after Kepler is direct imaging combined with spectroscopic characterization of exo-Earths, which would tell us whether they possess an atmosphere, oxygen, liquid water, as well as other biomarkers.  Direct-imaging requires extremely performant coronagraphic and wavefront control techniques. Although these tools are getting more mature for ground-based telescopes, they are still in development for space-based missions. 
Moreover, direct imaging of planets is still limited to single star systems and to planets found in a small field of view around the parent star. The first limitation is due to the fact that the light coming from the companion star pollutes the region around the main target. The second limitation is due to the number of degree of freedom of the deformable mirror used to create a dark zone.

We identified three main challenges associated with double-star (or multi-star) systems coronagraphy.
\begin{enumerate}
\item Often, multi-star separation is typically beyond the outer working angle (Half-Nyquist frequency) of the deformable mirror and therefore, the light coming from the companion star is difficult to suppress. 
\item Secondly, it is challenging to separate and independently remove overlapping speckles from multiple stars. 
\item Finally, demonstrating the first two points in broadband light will add to the challenge. 
\end{enumerate}

The main impact of this work is to enable direct imaging of planetary systems and disks around multiple star systems as well as in regions far from the star. This can be done at little additional hardware cost or changes to existing mission concepts, such as AFTA, Exo-C, Exo-S, and EXCEDE to name a few (coming online after 2020). This will greatly multiply the science yield of these missions. In addition, this solution enables a large space telescope to survey multi-star systems, magnifying its science return significantly.
One obvious target would be our nearest-neighbor star, Alpha Centauri ($\alpha$ Cen). It does also potentially enable the detection of biomarkers on Earth-like planets (if they exist) around $\alpha$ Cen, with a small and cheap space telescope of 25 cm diameter or even a balloon, potentially decades sooner than a large space coronagraph could do the same around a single star system. The next easiest stars are about three times more challenging considering that stars of comparable proximity to $\alpha$Cen are all very dim, and stars of comparable brightness are about three times farther away.

In this paper, we discuss the first challenge regarding the outer working angle. We demonstrate a new method called Aliased Wavefront Control (AWC), for which a mild grating (or an existing pattern commonly found on many Deformable Mirrors (DMs) and Micro-Electro-Mechanical Systems (MEMS) left over from their manufacturing process) effectively aliases low-spatial frequency modes of the DM into higher frequencies, enabling the DM to remove speckles well beyond the DMÕs Nyquist frequency. In effect, aliasing is used as a feature rather than a limitation.

 The second section explains the aliased wavefront sensing followed in section 3 by  simulation results. Finally, the last section shows options to create the diffracted dots needed for the method to work. The results are shown in the context of the EXO-C mission for which the telescope diameter is 1.5m and the time of observation would be around 2025.

%%%%%%%%%%%%%%%%%%%%%%%%%%%%%%%%%%%%%%%%%%%
\section{Aliased Wavefront Control}
Let's consider an N$\times$N DM. With such a DM, the limit of the correctable region is defined by the Nyquist frequency $N/2*\lambda/d$. To correct past this frequency, and create a dark hole at more than $N/2*\lambda/d$, simulations show that a diffractive grid is needed. A mild grating or an existing pattern commonly found on many DMs left over from their manufacturing process  can be used to create a controlled diffracted image of the off-axis star close to the on-axis star, placing the diffracted off-axis image within the DM control region. This grating can be placed in any pupil plane. This enables removing the speckles from both stars. This solution is innovative because it uses the DM in a regime outside its nominal control range, where the DM experiences what is typically a harmful side effect: spatial aliasing. We use aliasing as a feature (inspired by signal processing theory) and call this method ÒAliased Wavefront ControlÓ, or ACW.

In this paper, we present the simulation results of the preliminary proof of concept. The process is to simulate a science target observed with a system composed of a DM on axis and demonstrate that it is possible to create a dark zone past the Nyquist frequency of such a DM.  As mentioned in the introduction, we are interested in simulating the observation of the components of $\alpha$ Cen with a 1.5m telescope in monochromatic and polychromatic light (3\% bandwidth with a central wavelength of 770nm). The expected separation of $\alpha$ Cen components A and B in 2025 is about 10 arcsec, which corresponds to about 100 $\lambda/d$ away at a wavelength of 770 nm. For a potential more modest 25cm space telescope size, such a separation would correspond to 17 $\lambda/d$ and be at the edge of a 32x32 MEMS, making it still super Nyquist.

Since this paper is focusing on the feasibility of the method, we simplified the simulations to a simple telescope observing the un-occulted companion of the main target located at 100 $\lambda/D$. The goal is to demonstrate that it is indeed possible to cancel the light after the Nyquist frequency of a DM in this configuration.

\begin{figure}[h]
\begin{center}
\begin{tabular}{ccc}
\includegraphics[scale=0.3]{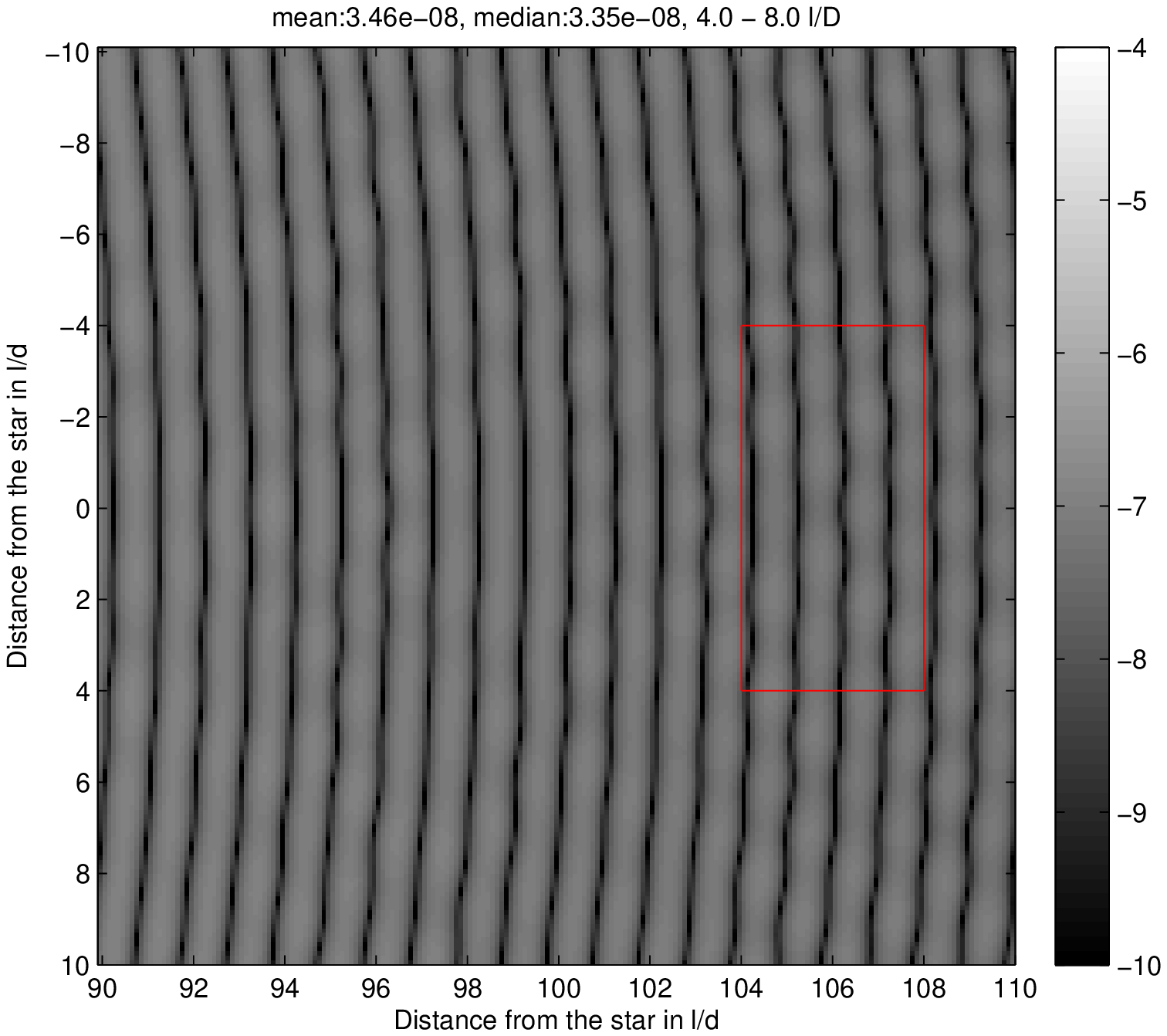}& \includegraphics[scale=0.3]{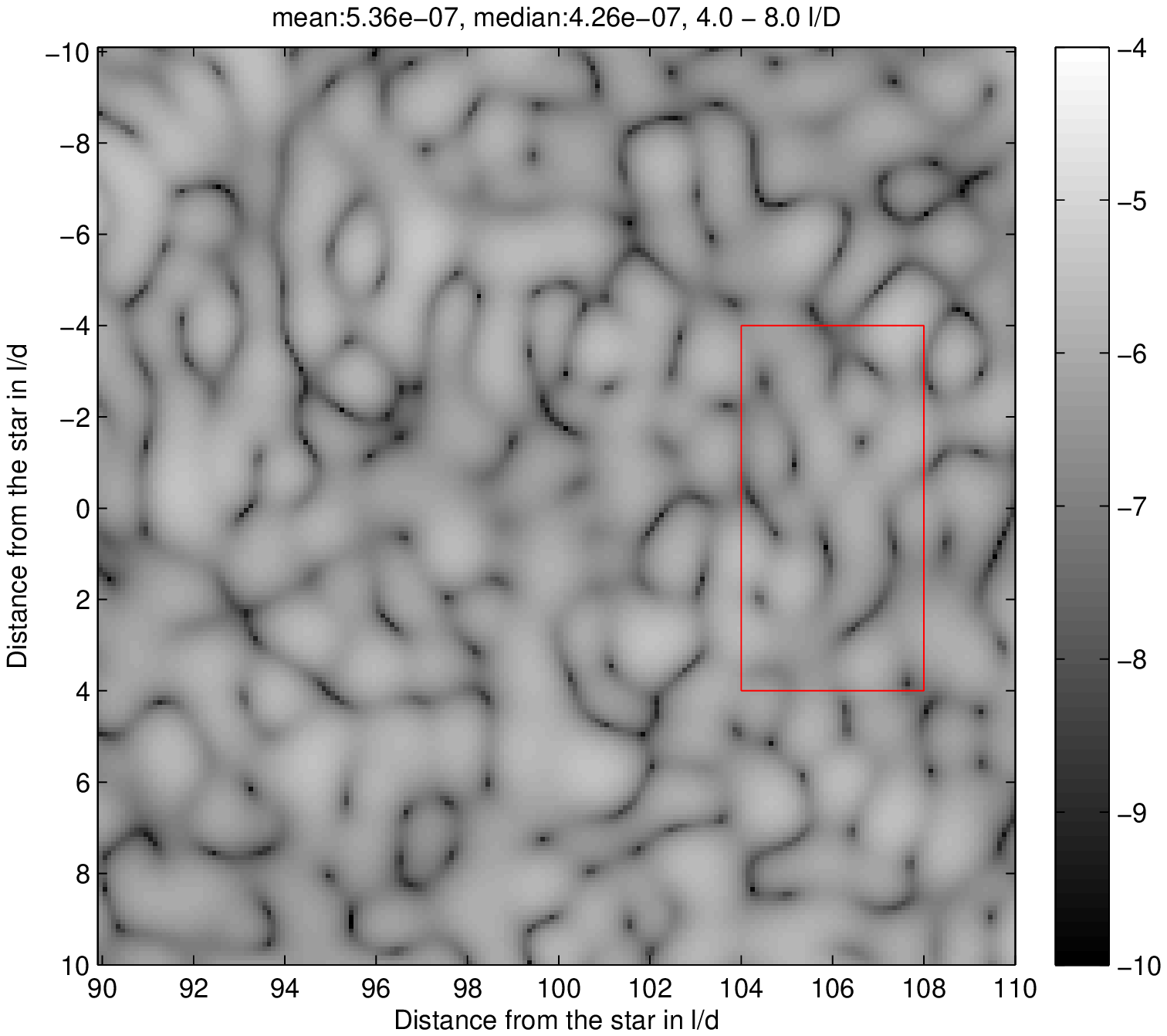}\\
\end{tabular}
\caption{Light diffracted at 100 $\lambda/D$ in monochromatic light , without aberration (left) and with 25nm rms aberrations (right). The size of the region of interest is 4$\times$8 $\lambda/D$. The median contrast intensity of the order of 4e$^{-8}$ without aberrations and 4e$^{-7}$ with 25nm rms of aberrations. The contrast is shown is in a single star case scenario.}
\label{fig:DiffLight}
\end{center}
\end{figure}

\begin{figure}[h]
\begin{center}
\includegraphics[scale=0.3]{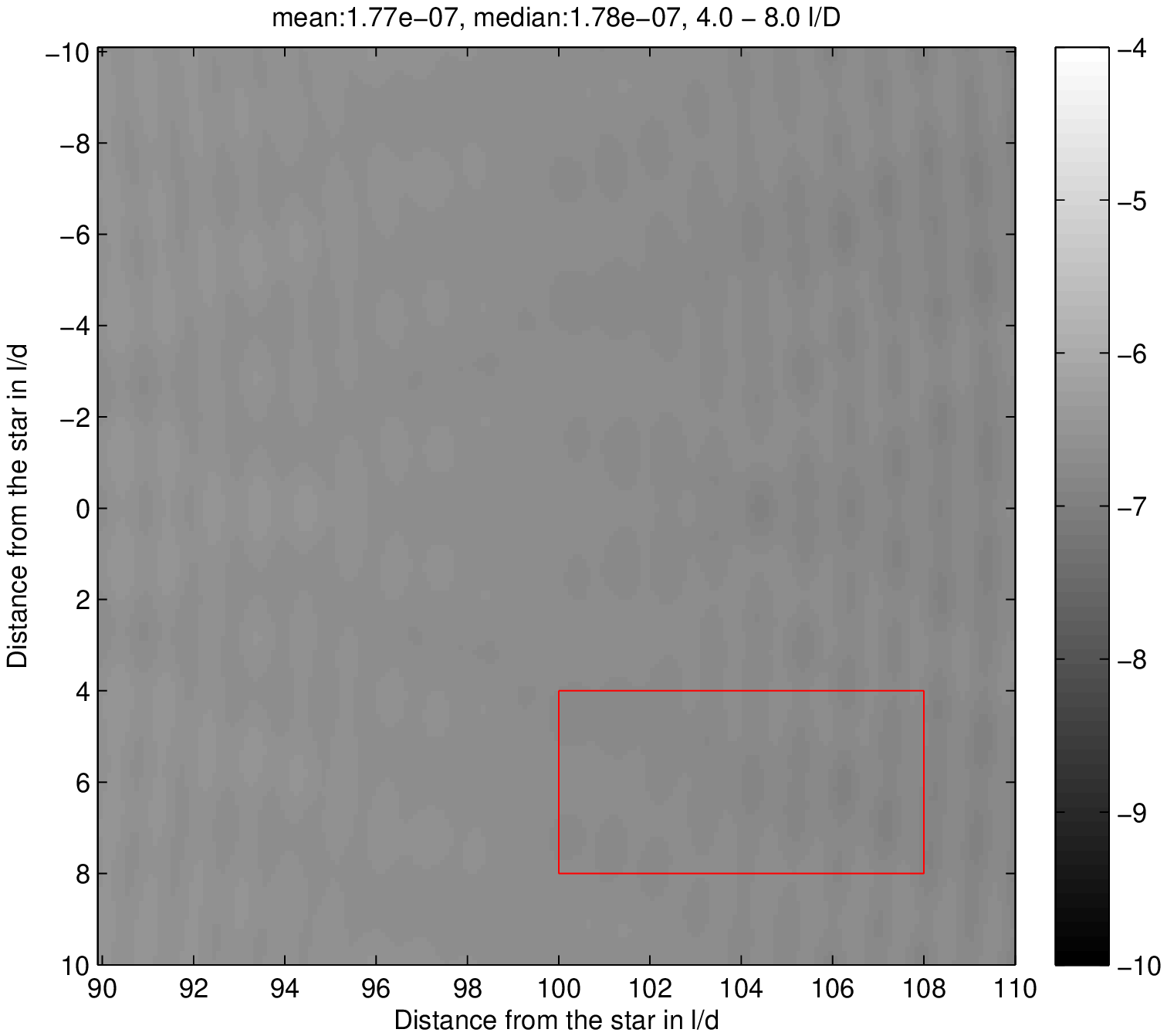}
 \includegraphics[scale=0.3]{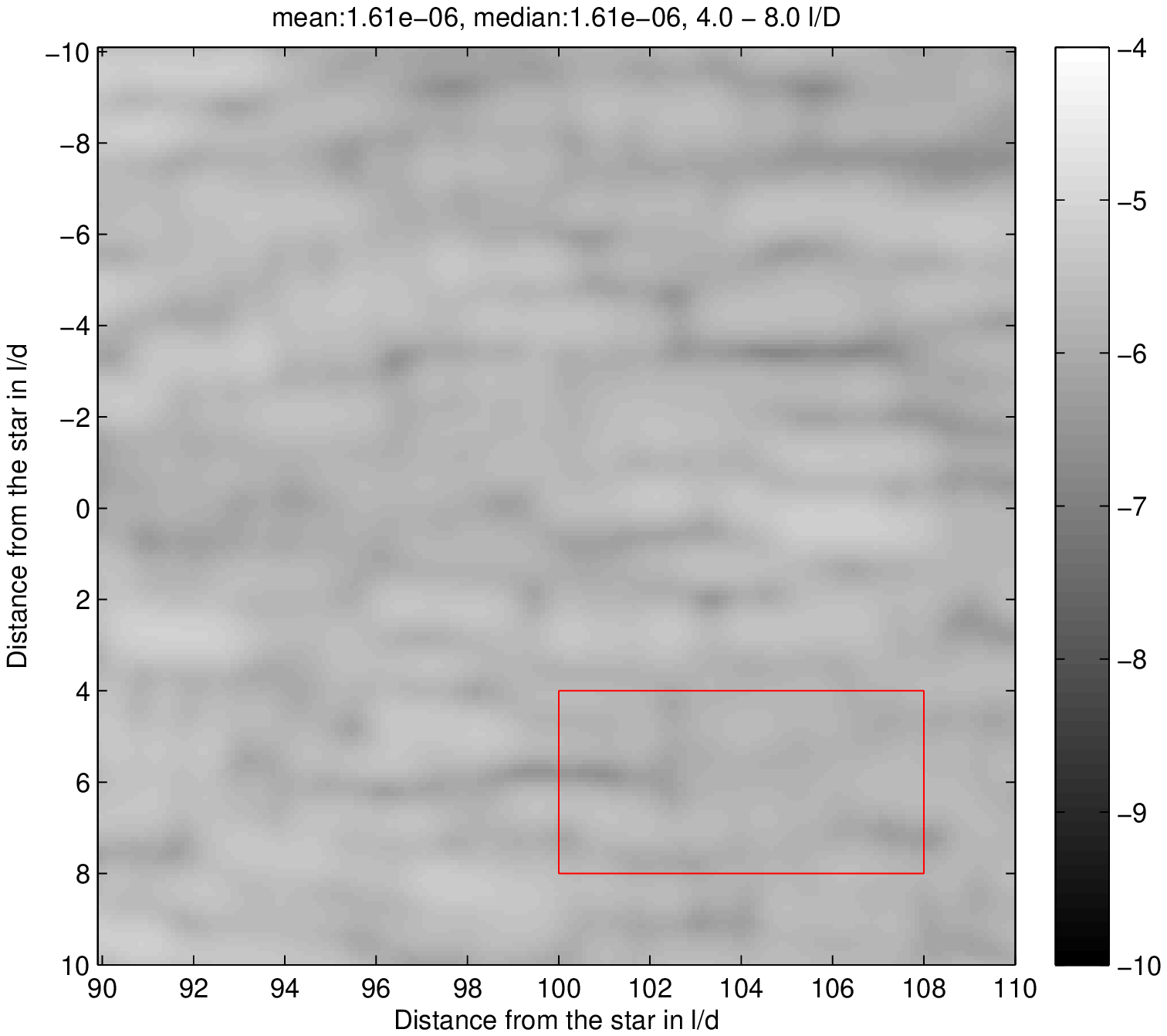}
\caption{ Light diffracted past 100 $\lambda/D$ in polychromatic light (3\% bandwidth), without aberration (left) and with 25nm rms aberrations (right). The size of the region of interest is 4$\times$8 $\lambda/D$. The median contrast intensity of the order of 2e$^{-7}$ of without aberrations and 2e$^{-6}$ with 20nm rms of aberrations}
\label{fig:DiffLightPoly}
\end{center}
\end{figure}

As mentioned above, we are considering both stars as potential planet host. The achieved contrast is first calculated relative to the star in a single-star case and then depending on if the focus is A or B, one has to multiple or divide by the ratio of the intensity of the binary star. If not otherwise specified, contrasts in the rest of the paper are quoted relative to the originating star in the single case scenario. Also let's call the main target of interest "the target" and the contaminating star "the companion".
Figures \ref{fig:DiffLight} and \ref{fig:DiffLightPoly} show the amount of diffracted light coming from the companion in $\alpha$ Cen in the region of interest of the second component of $\alpha$ Cen, both in mono and poly chromatic light and without any diffractive grating. The figures also show the goal for the dark zone region as a red box. We use these regions as a starting point for the median amount of light diffracted (see following section). 

The amount of light originating from the companion of the target of interest depends on the amount of aberrations in the system. In monochromatic light and in presence of  no aberrations, the amount of light diffracted in the region of interest is of the order of 4e$^{-8}$.
In the case of $\alpha$ Cen the magnitude of the two components are respectively 0 and 1.3 for A and B. If B is the main target of interest, the leakage from A would become three times worse, leading to 1.2e$^{-7}$ and if A is the main target the contrast gets better reaching 1.33e$^{-8}$.  Following the same logic in presence of 25nm rms aberrations, one gets 4e$^{-7}$ contrast if taking the A as a reference or 1.2e$^{-6}$ if taking B as a reference.

As expected, in polychromatic light, the leakage leads to a worse contrast. Without any aberration, the contrast obtained is 2e$^{-7}$ if A is the reference and 6e$^{-7}$ if B is the reference. With 20nm rms of aberrations, these numbers become 2e$^{- 6}$ if A is the reference and 6e$^{-6}$ if B is the reference.

%%%%%%%%%%%%%%%%%%%%%%%%%%%%%%%%%%%%%%%%%%%

\section{Simulation results}

\subsection{Method}
In order to create a dark zone around one of the two components of the binary, we consider the correction part of an Electric Field Conjugation algorithm (EFC) \cite{Giveon07} coded in Matlab. We assume that we know the wavefront to be corrected and therefore do not consider the estimation part of EFC. Also we simulate a portion of an Airy pattern resulting from a normal circular aperture at 100 $\lambda$/d. 
To be able to correct past the Nyquist frequency, simulation showed that it is required to add a grid of dots in a pupil plane to create the wanted dark zone.  To create the grid of dots, we used a mask of uniform intensity equal to 1 with a grid of 0.  The dots are the size of a pixel and therefore depend on the resolution of the simulation. We will not study the impact of the resolution of these dots on the performance in this paper. The period of the grid is set such that there will be a spike next to the zone we would like to correct. Depending on separation of the two components of the binary, one can use the regular pattern that MEMS (Micro-Electro-Mechanical system) create. Because they have a fixed number of actuators the frequency that the MEMS can create has to be a multiple of the number of actuators (here 32). Since 100 $\lambda$/d is not a multiple of 32 (common number of actuators across a MEMS at the ACE laboratory) we need to add a grid on top of the MEMS in intensity, The choice of the frequency of the dots of the grid is set up by the intensity of the light diffracted by the grid. We chose it such that the median contrast with the grid is close to the one without the grid. For the particular case, the resulting frequency was 50 cycles per aperture. This means that the diffracted dots seen in the images in the rest of the paper corresponds to the second diffractive order.

Another parameter that can be adjusted in order to control the performance is the width of the dark zone region. Indeed the bigger the region, the more DM stroke is needed to achieve deeper contrast. We used a 4$\times$8 $\lambda/d$ region, which is a good compromise between the performance and the discovery region. 

The simulations were done both in monochromatic light (770 nm) and in polychromatic light. From the polychromatic light, we are at this point limited to 3\% bandwidth because of computer memory instead of feasibility of the method. We are in the process of reimplementing the code to free up memory.
Finally, we also studied the case of a non-aberrated and aberrated wavefront.  The aberrations were introduced in the pupil plane as a power law with a coefficient equal to -2.

\subsection{Results}
\begin{figure}[h]
\begin{center}
\begin{tabular}{cc}
\includegraphics[scale=0.4]{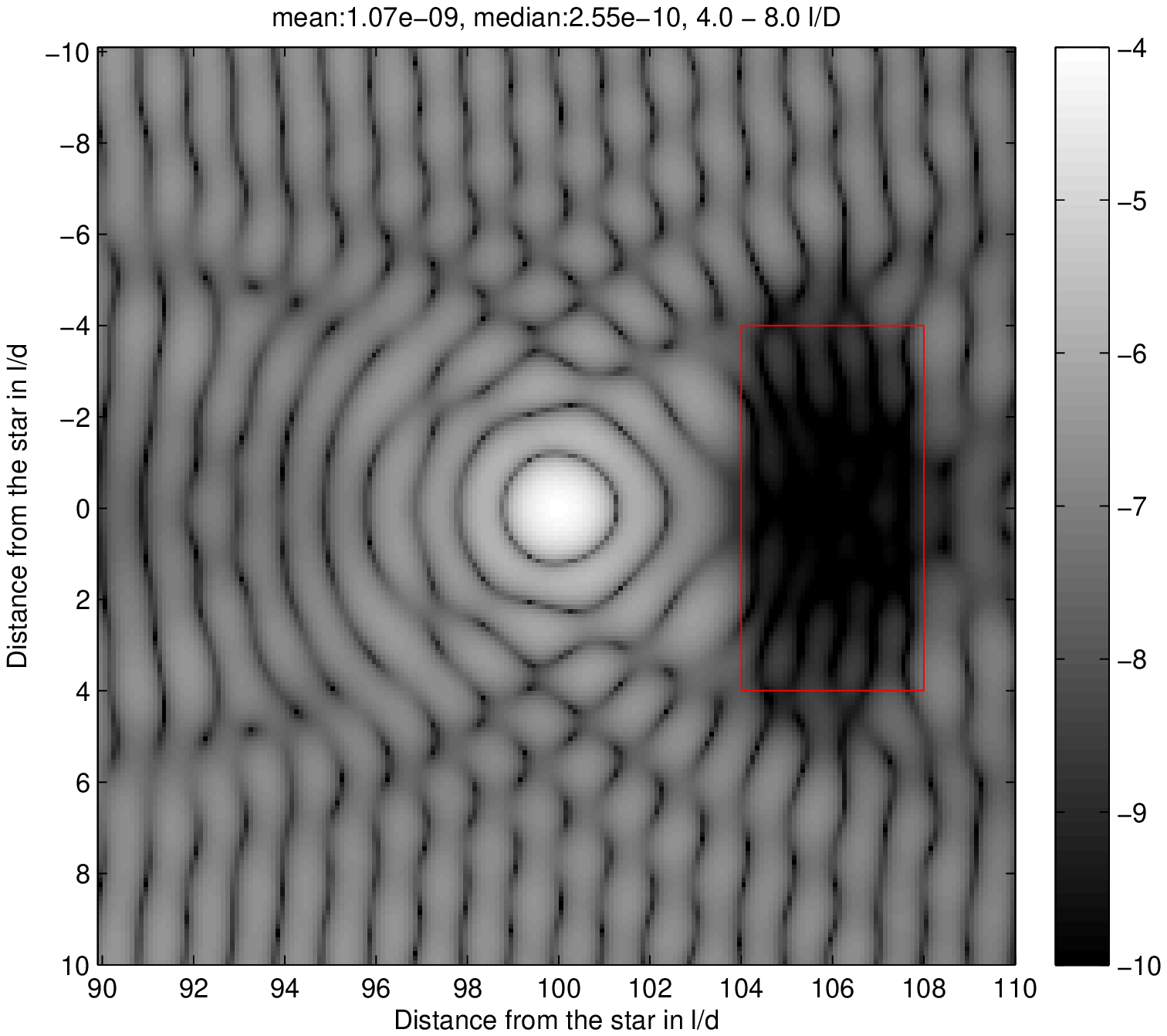}& \includegraphics[scale=0.4]{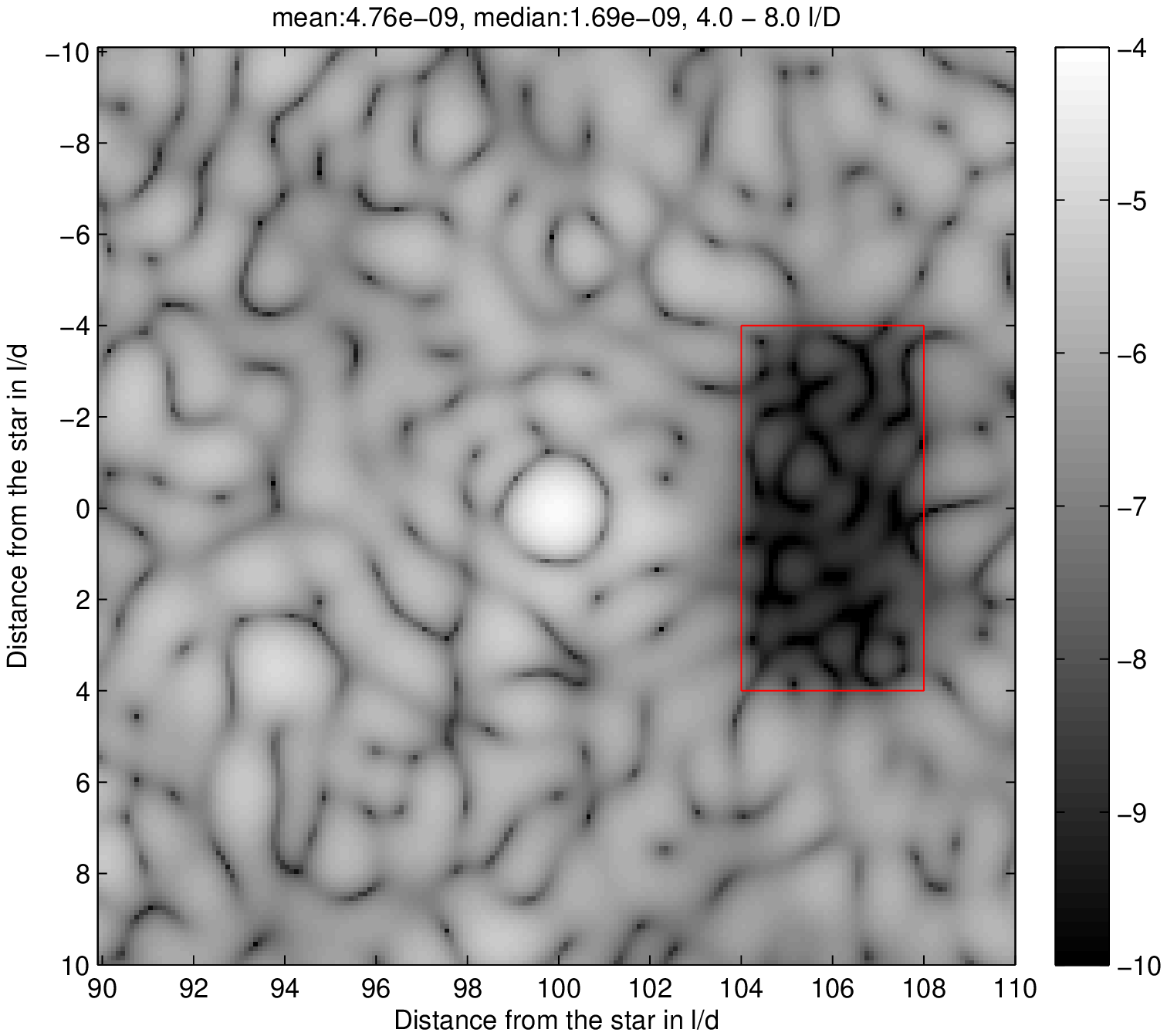}\\
\end{tabular}
\caption{Monochromatic case, 4x8 $\lambda/D$, Left: 0nm rms aberrations, 200 iterations.  Right: 25nm rms aberrations,  200 iterations. }
\label{fig:monolightres}
\end{center}
\end{figure}

\paragraph{Monochromatic Light}
In monochromatic light, the grid introduced creates dots at 100 $\lambda$/d with an intensity of 1.34e$^{-3}$ relative to the central star.  Figure \ref{fig:monolightres} shows the results after 200 iterations without aberrations (left) and with 25nm rms of aberrations. The median contrast obtained without any aberrations is 2.5$e^{-10}$ and with aberration we reach a contrast of 1.7$e^{-9}$. This corresponds to a factor 100 improvement from the no grid simulation for both the aberrated and non-aberrated cases. In a follow up paper, this will be optimized to be able to reach contrast levels of the order of 1$e^{-10}$, contrast required to detect earth like planets.

\begin{figure}[h]
\begin{center}
\begin{tabular}{cc}
\includegraphics[scale=0.4]{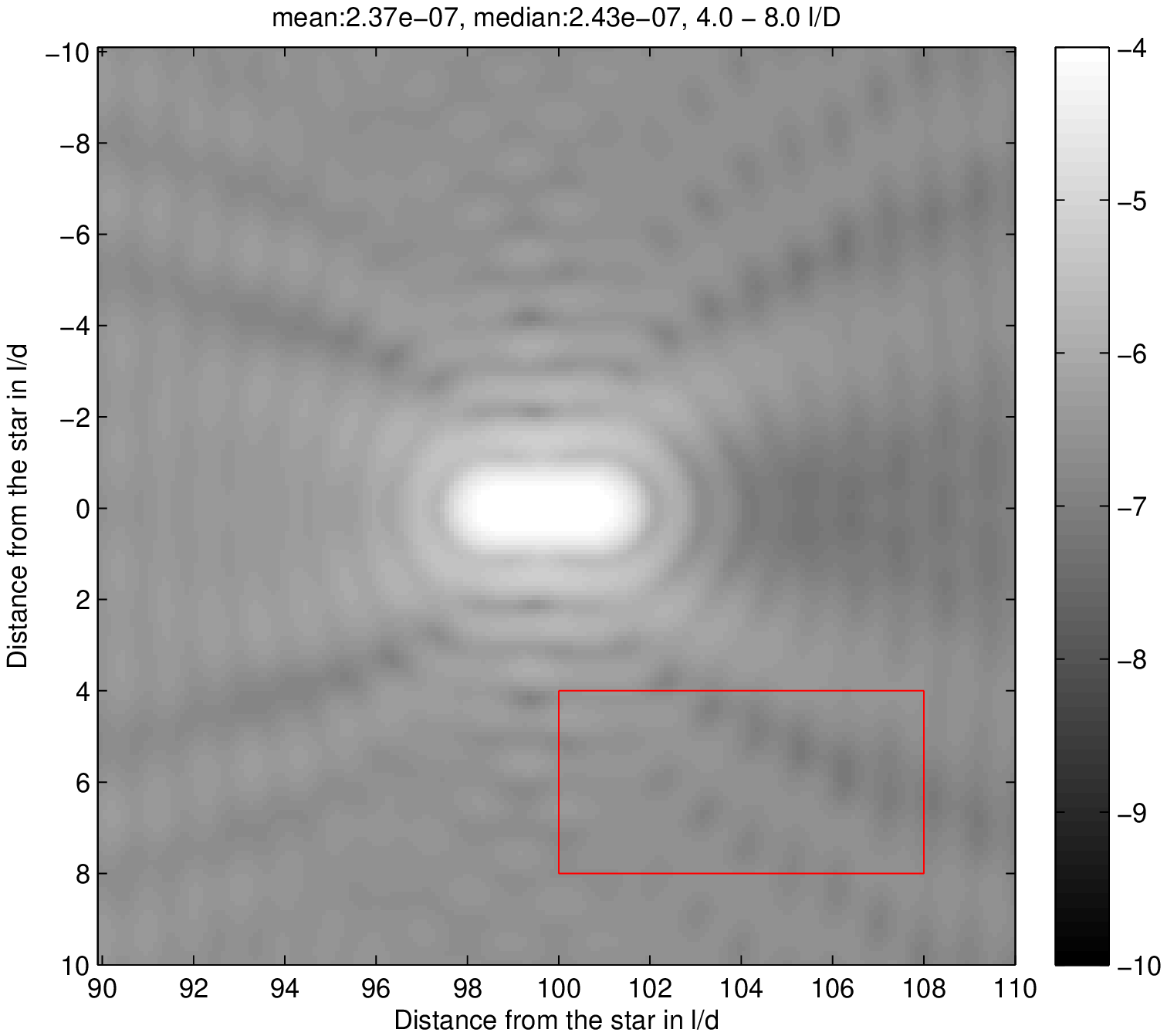}& \includegraphics[scale=0.4]{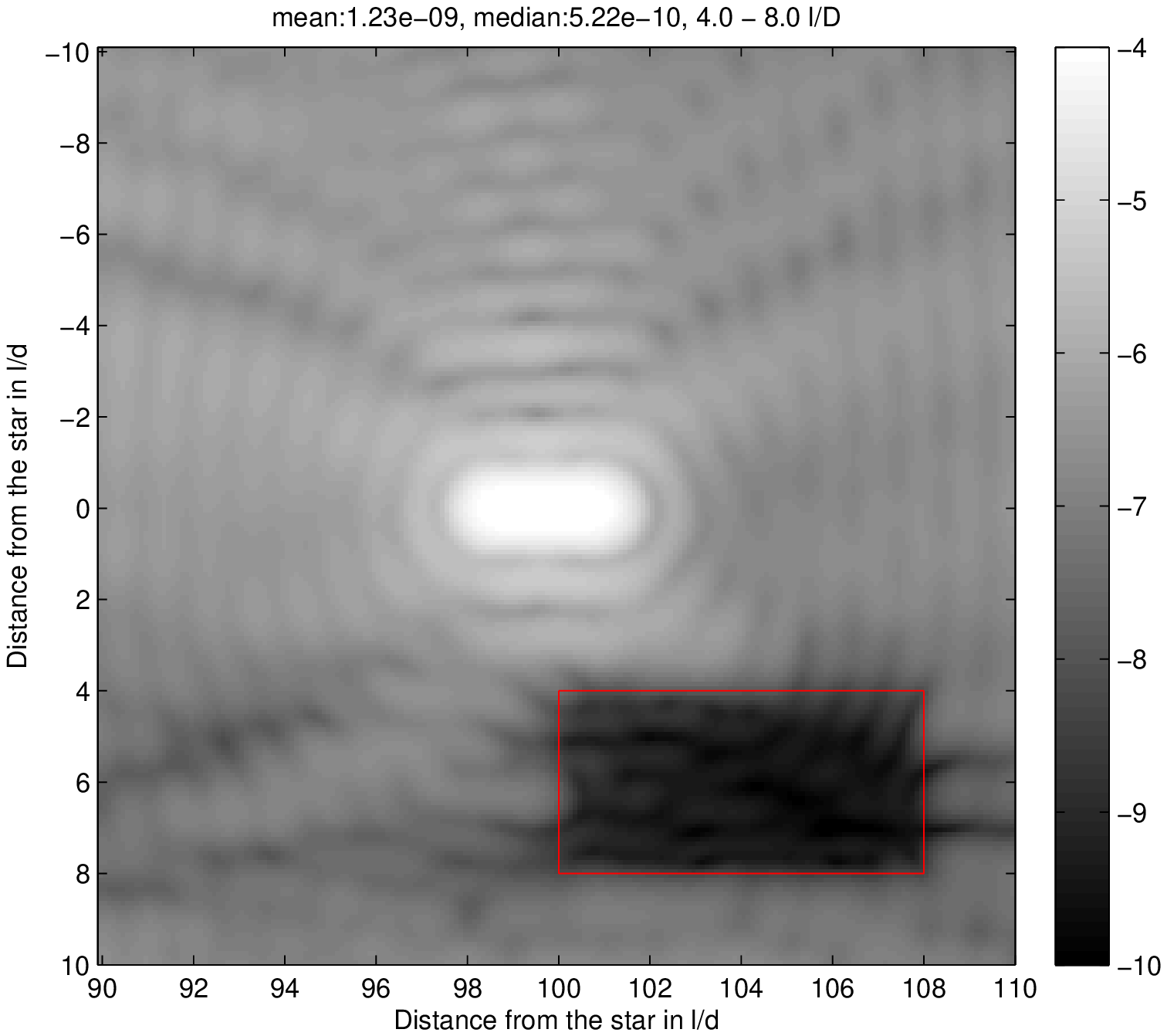}\\
\includegraphics[scale=0.4]{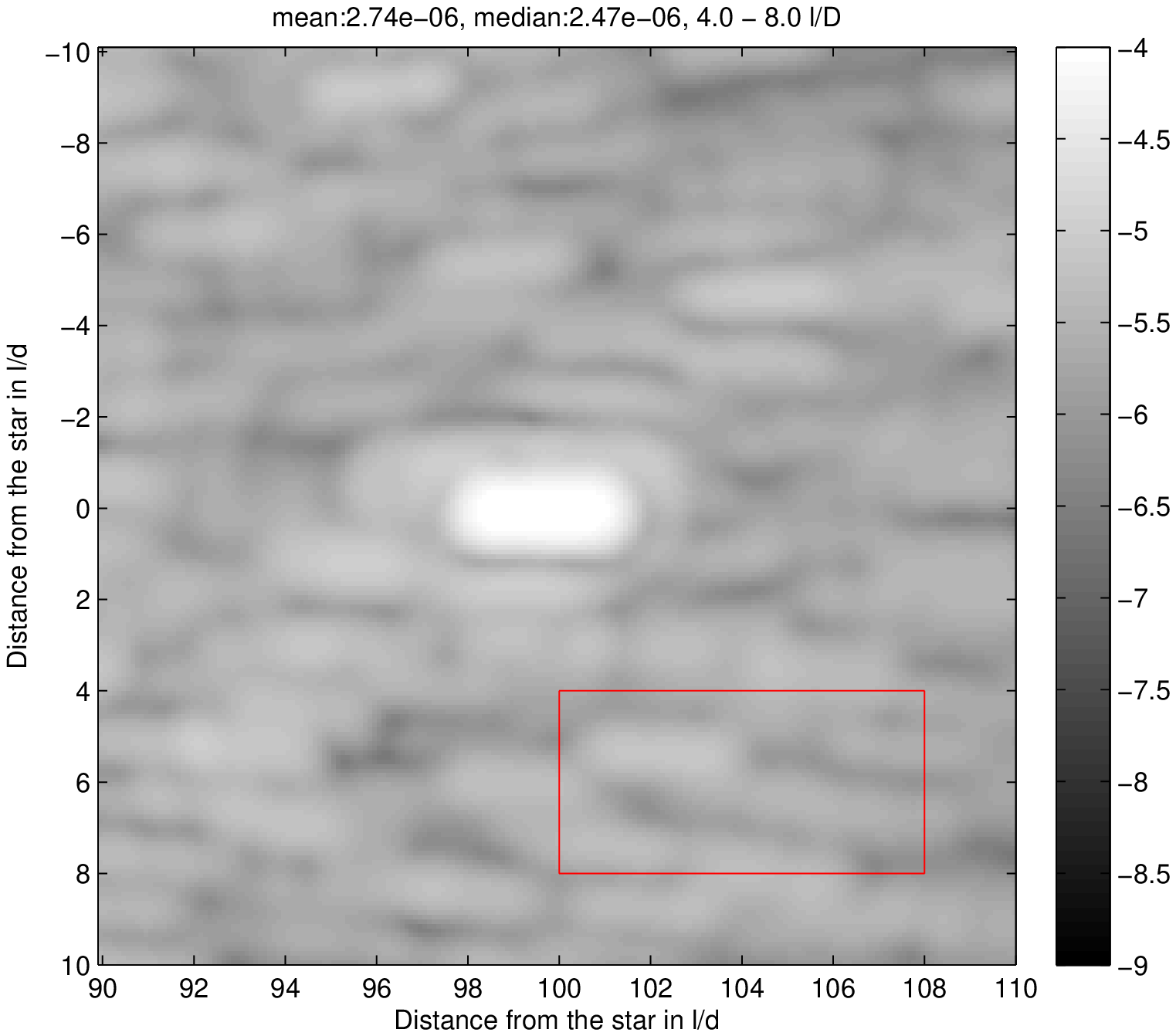}& \includegraphics[scale=0.4]{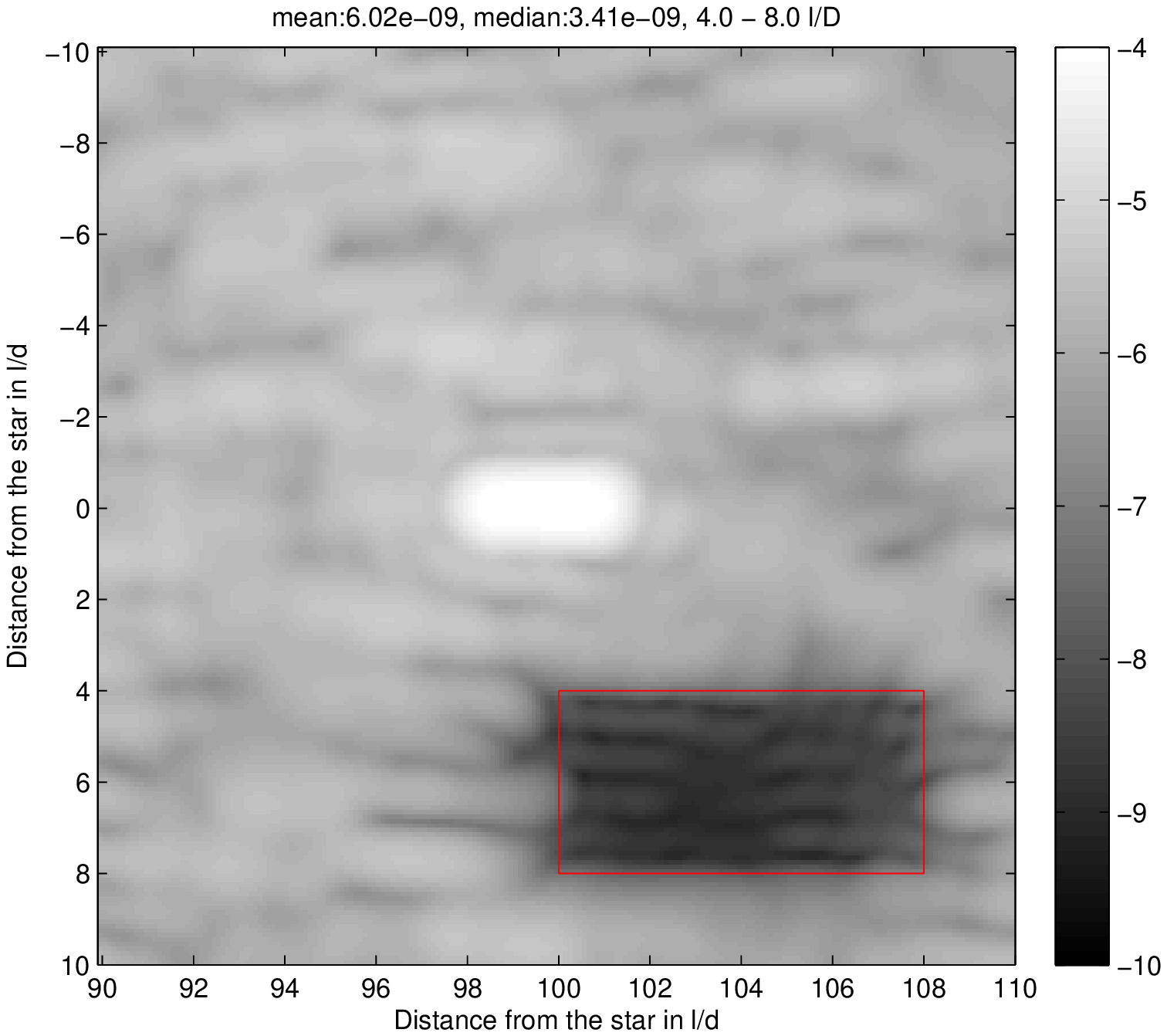}\\
\end{tabular}
\caption{ Polychromatic case, 4-8 $\lambda/D$, 1) 0 nm rms, 200 iterations  2) 25nmrms, 300 iterations}
\label{fig:Poly3}
\end{center}
\end{figure}

\paragraph{Polychromatic Light}
More realistic simulations include in polychromatic light. Therefore it is useful to know what performance we can achieve in broadband light. In this section, we will show corrections over a 3\% bandwidth. We calculate the G matrix for a discrete number of wavelength within the 3\% bandwidth \cite{Giveon07}. In order to get an accurate image in polychromatic light, we need a wavelength sampling sufficient to create a smooth elongated spot. For a 3\% bandwidth, we estimated 5 points over the bandwidth to be a good compromise between computational speed and resolution.  
Figure \ref{fig:Poly3} shows the results without and with aberrations (20nm rms).  The median contrast obtained without any aberrations is 5.2$e^{-10}$ . With aberrations, we reach 3.4$e^{-9}$. This corresponds to almost a factor 100 from the no grid simulation for both aberrated and non aberrated cases. 

Table \ref{tab:res} shows the summary of the different simulation results with respect to either the target, or the companion. We consider both cases for which the target is A or B.

\begin{table}[h]
\begin{center}
\caption{\label{tab:res} Summary of performance achieved with the simulations.}
\begin{tabular}{lcccccc}
 & Aberration & Self reference & A ref & B ref \\
\hline
No Grid,  Monochromatic, No Correction       & 0nm & 3.5e-8 & 1.2e-8&  1.1e-7\\
No Grid, Polychromatic(3\%), No Correction & 0nm & 1.8e-7  & 6.0e-8& 5.4e-7\\
Grid, Monochromatic, Correction                   &0nm&2.6e-10  & 8.7e-10&7e-10\\
Grid, Polychromatic(3\%), Correction             &0nm&5.2e-10 & 1.7e-10&1.6e-9\\
\hline
No Grid,  Monochromatic, No Correction       & 25 nm& 5.4e-7&1.8e-7&1.6e-6\\
No Grid, Polychromatic (3\%), No Correction & 20nm & 1.6e-6&5.3e-7&4.8e-6\\
Grid, Monochromatic, Correction                     &25nm&4.8e-9&1.6e-9& 1.4e-8\\
Grid, Polychromatic(3\%), Correction              &20nm& 6.0e-9&2e-9&1.8e-8\\
\hline
\end{tabular}
\end{center}
\end{table}

%%%%%%%%%%%%%%%%%%%%%%%%%%%%%%%%%%%%%%%%%%%
\section{Diffractive pupil design}
\begin{figure}[h]
\begin{center}
\includegraphics[scale=0.75]{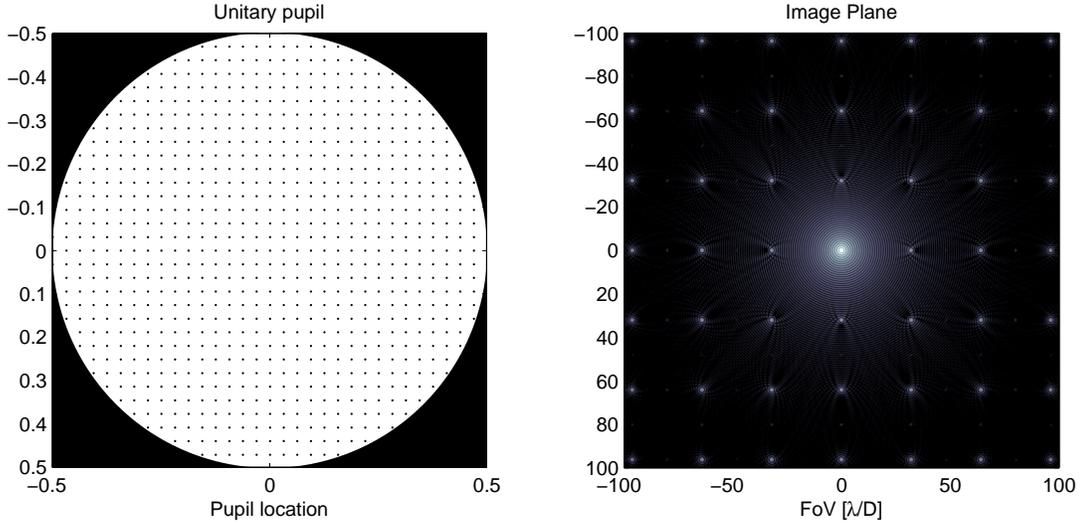}\\
\caption{\label{fig:Grid}  An example of the BMC Kilo DM that has 32 actuators across the pupil is shown on the left image. Each actuator has a small notch, which is shown as a dark dot. The image on the right shows a simulation of the image plane where we have PSF replicas every 32$\lambda/D$. This can also be a grid in a pupil plane.}
\end{center}
\end{figure}

As mentioned earlier, to execute the Aliased Waveform Control (AWC) algorithm it is necessary to create a replica of the target star in the location of interest. A diffractive pattern (DP) on the pupil plane can be used to create this replica. The two main design variables are the star replicaÕs location and intensity. The first variable drives the periodicity and geometry of the dot pattern in the pupil. The second variable controls the dot diameter \cite{Bendek13}. Under monochromatic illumination, the DP creates an inverse spatial frequency array of diffractive spots on the image plane. The dots in the pupil can be created using different technologies. In fact, as mentioned above, some DMs have a microscopic dot on top of each actuator due to manufacturing requirements. This creates a square grid of dots on the pupil resulting in an inverse spatial frequency array of diffracted replicas of the target star in the image plane as shown in Figure \ref{fig:Grid}. 

For a DM that will be placed on pupil plane of diameter D and has n number of actuators across the pupil arranged on square grid, the location of diffractive dots can be modeled as:

\begin{equation}
g(x,y)=A \times comb \left(\frac{xD}{n},\frac{xD}{n}\right),
\end{equation}

Where the comb function is a two-dimensional array of delta functions, and A represents the scaling factors required to maintain the normalization of delta functions, Note that the axis coordinates x and y have been multiplied by the aperture D to normalize the result to the aperture size. At the image plane the Fourier transform of this grid is obtained as,

\begin{equation}
G=F_{\xi}F_{\eta}[g(x,y)]=comb\left( \frac{n}{D}\xi, \frac{n}{D}\eta \right),
\end{equation}

Where $\xi$ and $\eta$ represents the transform variables and axes in the image plane. On the image plane we obtain a bi-dimensional comb function. The Fraunhofer far field imaging relation states that the transform variables $\xi$ and $\eta$ have a 1/$\lambda$-scaling factor resulting in replication of the PSF on square grid with spacing, 

\begin{equation}
\xi=\eta = \frac{n\lambda}{D},
\end{equation}

\begin{figure}[h]
\begin{center}
\includegraphics[scale=0.5]{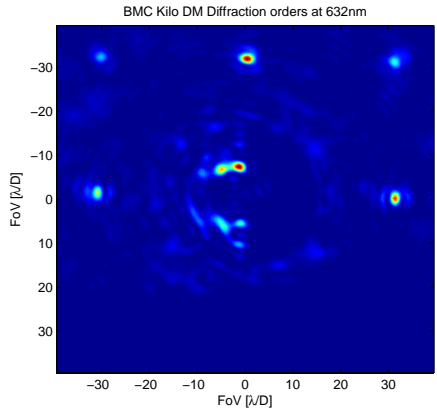}\\
\caption{\label{fig:BMC} Real image when the BMC Kilo DM has been installed in a reimaged pupil. We observed the PSF replicas every 32$\lambda/D$ as predicted by the diffractive theory.}
\end{center}
\end{figure}

For the particular case of the Boston Micro Machines Kilo DM that has 32 actuators across the pupil we obtain PSF replicas at 32$\lambda/D$, 64$\lambda/D$, 96$\lambda/D$ and other integer multiples. A real image obtained with a BMC Kilo DM is shown in \ref{fig:BMC}.

The pupil can be modeled as a convolution of the grid g with a circular function of diameter d that represents the dots, multiplied by a larger circular function of diameter D that represents the telescope aperture. The result on the image plane is a replica of the telescope PSF on every delta of the grid, which is modulated in intensity by an Airy function that has its first zero at 1.22$\lambda/d$ as a result of the dot size d. Normally the dot is very small creating a very wide modulation that slowly reduces the intensity of the dots as the FoV increases.

Note that the stability of the spikes depends on the mechanical stability of the dots on the pupil. Therefore, it is important to create the dots directly on the mirror coating for polychromatic light, the dots are replicated and scaled for each wavelength creating spikes. 

Using dots on the pupil has been proposed previously to calibrate dynamic distortions on wide-field optical systems enabling high-precision astrometric measurements \cite{Guyon12}. For this technique the dots can be arranged in the pupil using a hexagonal geometry allowing higher azimuthal sampling. The diffractive pupil spacing can be adjusted to create PSF replicas to run the AWC, and also to obtain high-precision astrometry on wide-field images. A description of the optimal hexagonal geometry for combining these two techniques has been published before \cite{Bendek13}.

\section{Conclusion}
In this paper, we showed that it is possible to create a dark zone past the Nyquist frequency of a DM using a diffractive grid in the pupil. This grid can be the DM itself or an additional mask. Performance was shown in the case of $\alpha$ Cen, an interesting target since it is the closest potential earth-like planet hosts. This can be applicable to star shade coronagraph for which we perfectly block the light coming from the parent star but still need to remove the light coming from the companion in the dark zone of interest.
The next step is to show that the DM can create a dark zone when adding both components. This will be done either using one or two DMs. The main impact of this work is to enable direct imaging of planetary systems and disks around multiple star systems as well as in regions far from the star. This can be done at little additional hardware cost or changes to existing mission concepts, such as AFTA, Exo-C, Exo-S, and EXCEDE to name a few, coming online after 2020. This will greatly multiply the science yield of these missions.

\acknowledgments
The material is based upon work supported by the National Aeronautics and Space Administration under Prime Contract Number NAS2-03144 awarded to the University of California, Santa Cruz, University Affiliated Research Center. This work was supported in part by the National Aeronautics and Space Administration's Ames Research Center, as well as the NASA Explorer program and the Technology Development for Exoplanet Missions (TDEM) program through solicitation NNH10ZDA001N-SAT at NASA's Science Mission Directorate. It was carried out at the NASA Ames Research Center. Any opinions, findings, and conclusions or recommendations expressed in this article are those of the authors and do not necessarily reflect the views of the National Aeronautics and Space Administration.

%\begin{thebibliography}{}
%\end{thebibliography}

%%%%%%%%%%%%%%%%%%%%%%%%%%%%%%%%%%%%%%%%%%%%%%%%%%%%%%%%%%%%%
%%%%% References %%%%%

\bibliography{SandrineThomasPaper}   %>>>> bibliography data in report.bib
\bibliographystyle{spiebib}   %>>>> makes bibtex use spiebib.bst

\end{document}